\documentclass[%
 reprint,
 groupedaddress,
 aps, prl,
 floatfix
]{revtex4-1}

\usepackage{graphicx}	% Include figure files
\usepackage{natbib}

\usepackage[colorinlistoftodos]{todonotes}

\begin{document}

\preprint{APS/123-QED}

\title{
Splitting the Raman beamsplitter
}

\author{Matt Jaffe}
  \email{mjaffe@berkeley.edu}
\author{Victoria Xu}%
%\affiliation{Department of Physics, University of California, Berkeley, California 94720, USA}
\author{Philipp Haslinger}%
\altaffiliation[Now at: ]{Technische Universit\"at Wien - Atominstitut, Stadionallee 2, 1020 Wien, Austria}
%\affiliation{Department of Physics, University of California, Berkeley, California 94720, USA}
\author{Holger M{\"u}ller}%
\altaffiliation[Also at: ]{Molecular Biophysics and Integrated Bioimaging, Lawrence Berkeley National Laboratory}
\affiliation{Department of Physics, University of California, Berkeley, California 94720, USA}

\author{Paul Hamilton}
\affiliation{%
 Department of Physics and Astronomy, University of California, Los Angeles, California 90095, USA}

\date{\today}

\begin{abstract}
We present an atom interferometry technique in which the beamsplitter is split into two separate operations. A microwave pulse first creates a spin-state superposition, before optical adiabatic passage spatially separates the arms of that superposition. Despite using a thermal atom sample in a small ($600 \, \mu$m) interferometry beam, this procedure delivers an efficiency of $99\%$ per $\hbar k$ of momentum separation. Utilizing this efficiency, we first demonstrate interferometry with up to $16\hbar k$ momentum splitting and free-fall limited interrogation times. We then realize a single-source gradiometer, in which two interferometers measuring a relative phase originate from the same atomic wavefunction. Finally, we demonstrate a resonant interferometer with over 100 adiabatic passages, and thus over $ 400 \hbar k$ total momentum transferred.
\end{abstract}

\maketitle

Atom interferometers (AIs) have been used for many purposes, such as measuring fundamental constants \cite{Rosi2014PrecisionAtoms, Fixler2007AtomGravity, Bouchendira2011NewElectrodynamics, Parker2018MeasurementModel}, testing fundamental laws of physics \cite{Schlippert2014QuantumFall, Zhou2015TestInterferometer}, and as inertial sensors \cite{Peters1999MeasurementAtoms, Gustavson2000RotationGyroscope, Hu2013DemonstrationGravimeter}. Next-generation AIs \cite{Kovachy2015QuantumScale} target fundamental physics measurements \cite{Hartwig2015TestingInterferometer}, spaceborne operation, precision sensing in the field \cite{Freier2016MobileStability, Farah2014UndergroundGravimeter, Wu2017MultiaxisTrap}, and gravitational wave detection \cite{Graham2016ResonantInterferometry, Yu2011GravitationalInterferometers}. Their operation relies on creating superpositions of coherent matter waves and manipulating their spatial trajectories. A phase difference $\Delta\phi$ accumulates between arms of the superposition, which can be inferred from the probability $P$ for an atom to exit the interferometer in one of the output ports, given by $P = \frac{1}{2}\left(1 -\cos(\Delta\phi) \right)$. Sensitive interferometry techniques involve many such manipulations, necessitating efficient atom optics. Examples include large momentum transfer for increased sensitivity \cite{Muller2008AtomSplitters, Kovachy2015QuantumScale,McDonald201380Interferometer}, or a resonant AI consisting of many loops \cite{Graham2016ResonantInterferometry}.

These operations are generally performed with laser pulses. However, intensity variation across an atomic sample, due to the Gaussian beam profile of the laser, limits pulse efficiency. As a result, realizing interferometer geometries with more than just a few pulses requires a thick laser beam, an extremely cold atomic sample, or both. Adiabatic rapid passage (ARP) offers high efficiency despite a varying laser intensity. However, it essentially trades a spread in efficiency for a spread in phase, making its application to AIs \cite{Kotru2015Large-AreaPassage, Kovachy2012Adiabatic-rapid-passageOptics} difficult.

In this Letter, we demonstrate a technique enabling the use of adiabatic passage for matter wave optics with up to 99\% efficiency per $\hbar k$ of momentum transfer. We present flexible interferometer geometries utilizing this technique, including (i) large momentum transfer (LMT), (ii) single-source gradiometry, and (iii) multi-loop (up to 51 loops) resonant AIs for ac signal detection. The atom source is simple, using only optical molasses and Raman sideband cooling. The technique uses Raman transitions, providing state-labeled output ports, yet is highly insensitive to ac Stark shifts. These capabilities are acquired by splitting the beamsplitter operation.

In this context, a beamsplitter serves two purposes: it generates superposition, and puts the arms of that superposition into relative spatial motion. These are usually performed simultaneously, but could be performed separately \cite{Wu2009GravityInterferometer, Kotru2015TimekeepingInterferometers}, as in \cite{Machluf2013CoherentChip} with magnetic beam splitters, \cite{ Mizrahi2013UltrafastAtom} with trapped ions, and \cite{Featonby1998Separated-PathInterferometer} for temperature measurement.

\begin{figure}
\centering\includegraphics[width = 3.38in]{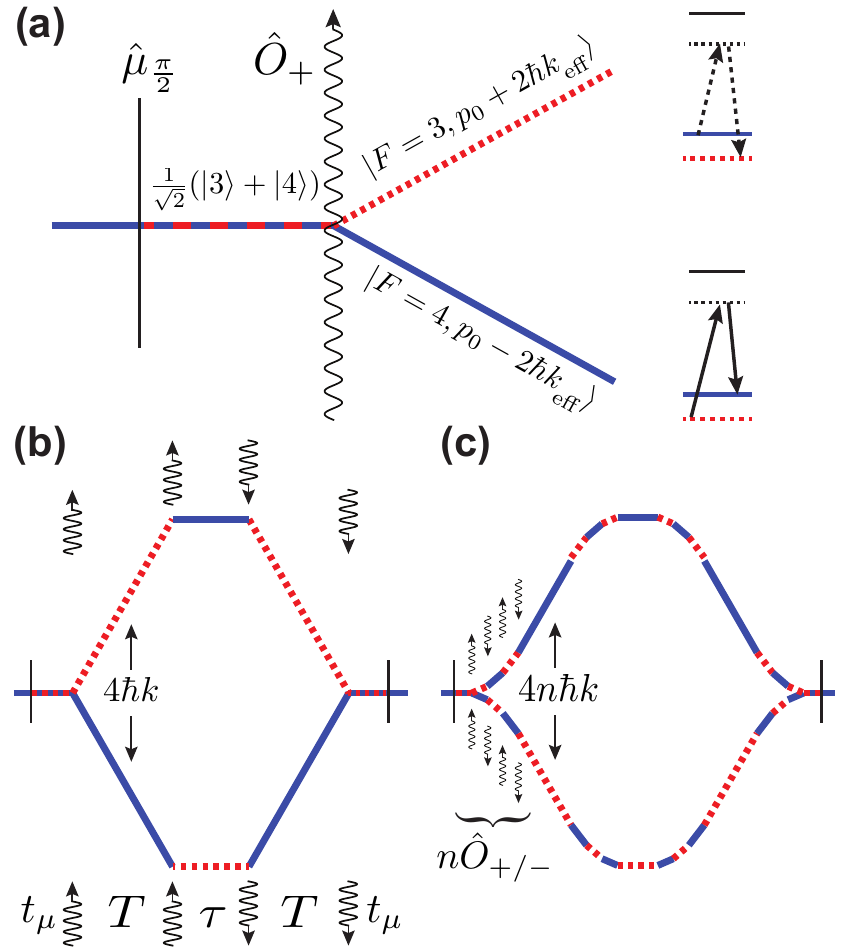} 
\caption{SDK interferometry. (a) SDK beamsplitter. A microwave $\frac{\pi}{2}$ pulse $\hat{\mu}_{\frac{\pi}{2}}$ puts the atom into a superposition of hyperfine states. A Raman adiabatic passage, $\hat{O}_{+}$, then delivers a spin-dependent kick to each arm of the superposition. The energy level diagrams at right show the transitions for both arms. (b) Basic SDK interferometer. During the wavepacket separation time $T$ the arms have $4\hbar k$ momentum separation, while $\tau$ denotes the time between halves of the SDK mirror pulse sequence, where the arms are at rest relative to each other.  (c) Large momentum transfer. Inverting the laser wavevectors kicks the arms in opposite directions, $\hat{O}_{-}$. Since both laser frequencies travel in both directions, either operation can be chosen (a large enough Doppler shift breaks the degeneracy).}
\label{sdk schematic}
\end{figure}

Here we demonstrate such a two-part beamsplitter that leverages the precision of the photon momentum for atom interferometry. Atom interferometers derive their accuracy from the fact that the photon momentum $\hbar \vec{k}$, which is given by the wavevector $\vec{k}$, determines the trajectories of the atoms precisely and thus provides a large and precisely known scale factor. Additionally, we perform each step of the beamsplitter efficiently: the superposition is generated using microwaves, and the spatial motion with optical adiabatic passage, see Fig. \ref{sdk schematic}(a). Since the direction of the interferometry kicks is determined by the initial state, we refer to it as a ``spin-dependent kick" (SDK), as inspired by the ion trapping scheme from \cite{Mizrahi2013UltrafastAtom, Campbell2017RotationIons}.

Our apparatus has been described previously \cite{Hamilton2015AtomCavity, Hamilton2015Atom-interferometryEnergy, Jaffe2017TestingMass}, and uses cesium atoms in the magnetically insensitive $m_F=0$ ground state, prepared by Raman sideband cooling in an optical lattice. Atoms are launched upwards using a frequency chirped pair of laser beams. The light used to manipulate the atoms uses an optical cavity, to provide mode cleaning and intensity build-up. Typical atom interferometers use large diameter beams to make the laser intensity as uniform as possible across the atomic cloud. Despite having only a 600 $\mu$m beam waist for an atom cloud of $\frac{1}{e}$ radius $\sim 350 \, \mu$m, we still achieve $>96\%$ efficiency per pulse.

Adiabatic passage provides independence of the Raman transition probability from the exact laser intensity, enabling high efficiency despite intensity variation over the sample. The atoms are driven by a pair of beams whose difference frequency is swept though Raman resonance, so that the state of the atom is adiabatically transferred from the initial to the final state. For the sweep, we use a cosine-squared temporal profile of the intensity, and thus the two-photon Rabi frequency 
$$ 
\Omega_{2\gamma} (t) = \Omega_0 \cos^{2}\left(\frac{\pi t}{2 \tau_{p}}\right),
$$ 
with $t\in[-\tau_{p},\tau_{p}]$ and $\tau_{p} = 100\,\mu$s. This pulse shape is used for its constant adiabaticity, which can be obtained by calculating the proper detuning $\delta(t)$ \cite{Bateman2007FractionalInterferometry}. This gives a bandwidth of $\Omega_{0}$ over which adiabatic passage transfers the atoms with a measured efficiency of $96\%$ ($\pm\sim$1\%, depending on the intensity used). Each pulse imparts $4 \hbar k$ momentum transfer, giving an efficiency of 99\% per $\hbar k$. For a 10\% overall efficiency, using SDK pulses increases the total possible momentum transfer by over an order of magnitude, from 12 $\hbar k$ (6 Raman pulses each 70\% efficient) to 260 $\hbar k$ (65 SDK pulses). This efficiency improvement is limited only by available laser power. Due to a fiber EOM damage threshold, only 12 mW are incident on the cavity at present.

Interferometers can be realized by combining SDKs. The simplest case (Fig. \ref{sdk schematic} (b)) is: one SDK beam splitter, followed by two adiabatic passages to invert the direction of the interferometer arms, and a final SDK beam splitter to combine the wave packets for interference. This interferometer has twice the momentum transfer of a traditional Raman interferometer.

We realize even higher momentum transfer by cascading SDKs as shown in Fig. \ref{sdk schematic} (c). Alternating between $\hat{O}_{+}$ and $\hat{O}_{-}$ pulses allows momentum transfer in the same direction as the spin state is toggled between $F=3$ and $F=4$. (This toggling could be avoided by inserting microwave $\pi$ pulses between the optical pulses, but this proved less efficient in our apparatus.) A $4n \hbar k$ interferometer, where $n=1,2,3,\ldots$, is realized by consecutive pulses to first accelerate the arms away from each other, then invert relative momentum, and finally recombine. The phase difference between the arms of this interferometer is given by
$$
\Delta\phi = 4 n \left(\vec{k} \cdot \vec{a}\right) T (T + \tau),
$$
where $\vec{a}$ is the acceleration experienced by the atom, and the times $T$, $\tau$ are defined in Fig. \ref{sdk schematic}.

\begin{figure}
\centering
\includegraphics[width = 3.38in]
{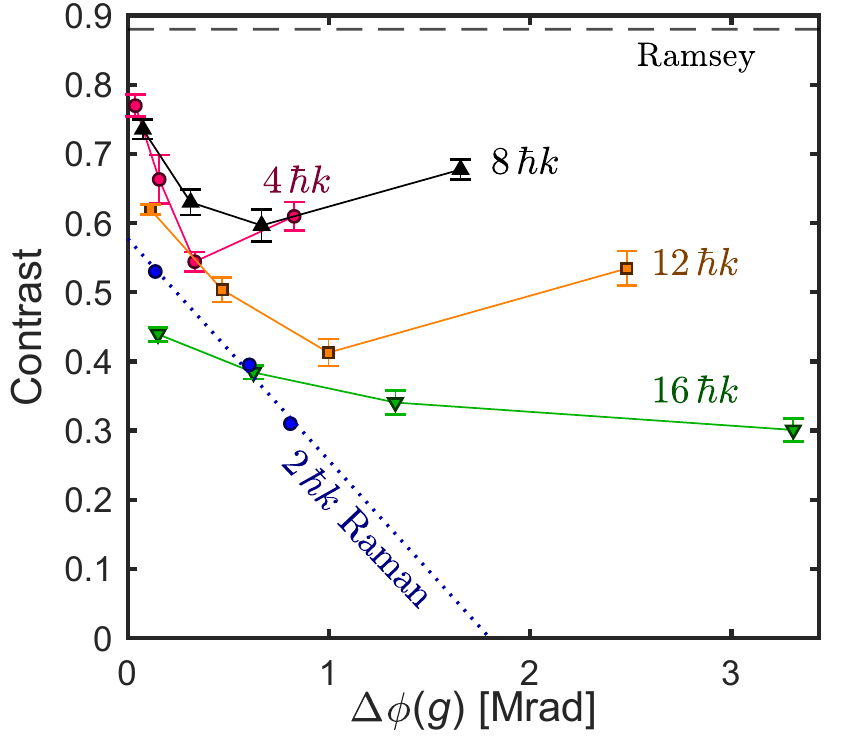} 
\caption{SDK interferometer contrast as a function of gravity phase $\Delta \phi$, measured for various orders of momentum transfer at wavepacket separation times $T$ = 5, 15, 25, and 44 ms. The gravity phase $\Delta \phi (g)$ is due to the acceleration from Earth's gravity, $g \approx 9.8\,\text{m}/\text{s}^{2}$. High visibility fringes are observed for $\Delta \phi \lesssim$ 0.5 Mrad, after which vibration noise dominates. Contrast is therefore determined by fitting histograms of $\sim200$ interferometer outputs to an arcsine probability distribution function. Error bars represent the 1$\sigma$ statistical uncertainty in the contrast fit parameter. The blue dotted line provides a comparison to traditional $2 \hbar k$ Raman interferometers in our apparatus with $T$ = 22, 55, and 65 ms.} 
\label{contrast decay}
\end{figure}

We have demonstrated time-of-flight limited performance for up to 16 $\hbar k$ momentum splitting (Fig. \ref{contrast decay}). The momentum separation in our current setup is limited by the use of the same laser frequencies to address both interferometer arms. As the separation increases, so does the relative Doppler shift between the arms until it exceeds the bandwidth over which the rapid adiabatic passage is efficient. This could be solved by higher laser power (allowing larger ARP bandwidth) or by toggling the Raman frequency to address the arms resonantly, one at a time.

Adiabatic passage can introduce a large spread in phase. To describe the origin of this phase spread, consider a two-level system on the Bloch sphere. In adiabatic passage, the state vector precesses around the drive vector with frequency $\Omega$, accumulating a large dynamical  phase $\gamma = \int \Omega(t')dt'$. When adiabatic passage is used to transfer between the poles of the Bloch sphere, this precession describes a narrow cone and therefore has a negligible effect on the final state. When applied to a superposition, however, the state starts and ends on the equator, with precession occurring in great circles on the Bloch sphere. Intensity variations now give rise to a spread (many $\pi$) in dynamic phases, dephasing the atom sample.

However, unlike an efficiency spread, a phase spread can be reversed. If two pulses are applied in quick succession with alternating sign of $\gamma$ (determined by initial state and sweep direction), the dynamic phase cancels. The contrast of our interferometer vanishes if such cancellation does not occur, for example if the sweep direction for a single ARP pulse is intentionally inverted. These de-/re-phasing effects have been explored using an atom interferometer with standard beamsplitters and ARP augmentation pulses in \cite{Kotru2015Large-AreaPassage}, though re-phasing imperfections limited pulse separation times to less than 10 ms.

We use several methods for effective rephasing and thus high contrast. First, we intensity-stabilize interferometry pulses to minimize optical power fluctuations. Second, we avoid large radial motion of the atoms by selectively detecting only the center of the atom cloud. The launch chirp is reversed to catch the atoms after the interferometer is closed. A low intensity is used so only the radially-centered portion of the cloud is caught (this is simplified in SDK interferometry, because both output ports have the same velocity). Third, because the intracavity intensity changes with frequency we adjust the input intensity of pulse pairs such that their Rabi frequencies are equal and thus their dynamic phases cancel. Finally, the largest source of rephasing errors in previous interferometers \cite{Kotru2015Large-AreaPassage} was the beam quality. In our apparatus, the optical cavity acts as a mode filter, providing very clean wavefronts.

As a result, we see excellent contrast out to $T = 44$\,ms (Fig. \ref{contrast decay}), limited only by the available free-fall time. SDK interferometers shown include a time $\tau \approx 20$ ms centered around the apex of the trajectory to avoid degeneracy between $\hat{O}_{+}$, $\hat{O}_{-}$ and the velocity-insensitive Raman transitions. The upper dashed line indicates the contrast of a Ramsey clock (i.e., only the $\hat{\mu}_{\frac{\pi}{2}}$ pulses) measured for various timings. Our interferometer with the largest scale factor ($16 \hbar k$, $T = 44$ ms, $\tau = 18$ ms) has a phase of $3.4 \times 10^{6}$ rad for the acceleration due to Earth's gravity. As a standard Raman Mach-Zehnder interferometer with the same $T = 44$ ms would have a phase of $0.28 \times 10^{6}$ radians, this represents over an order of magnitude improvement. 

Interestingly, the $8 \hbar k$-interferometer has higher contrast at long $T$ than the $4 \hbar k$-interferometer. For even-$n$, pulse pairs of a $4 n \hbar k$-interferometer can be separated in time by only the pulse duration $2\tau_{p}$. For odd-$n$, there is a pulse pair separated by $T$ (for us, up to 200 times longer). This gives more time for an atom to move within the laser beam profile, degrading dynamic phase cancellation.

This effect favors the long $T$ contrast of even-$n$ interferometers over odd-$n$. For $8 \hbar k$, a fit to an exponential decay of the contrast $c \propto \exp(-T/T_{0})$ gives a time constant $T_{0} = 260$\,ms, indicating scalability to even longer times. Finite adiabatic passage bandwidth hurts the contrast at larger momentum separation: at $12 \hbar k$ ($16 \hbar k$), the maximum Doppler shift between the arms is 100\,kHz (132\,kHz), while our SDK pulse bandwidth is 125\,kHz.

\begin{figure}
\centering
\includegraphics[width = 3.38in]{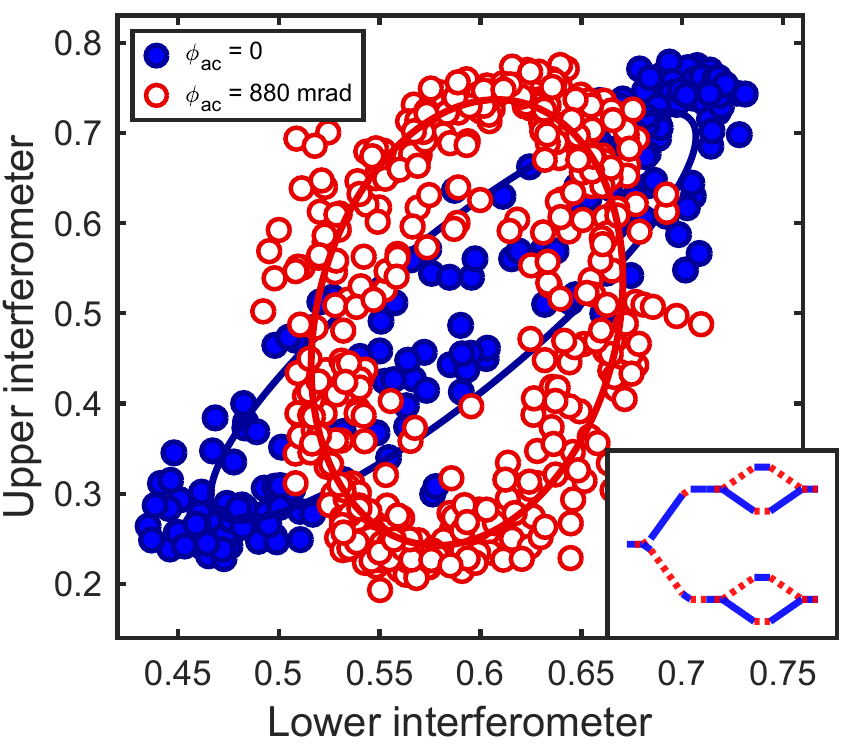} 
\caption{Single-source gradiometer. A schematic of the arm trajectories is shown in the inset. The first half of an $8 \, \hbar k$ SDK interferometer separates two arms. Once brought back to relative rest, the actual interferometer sequence begins, simultaneously addressing both arms. The phases of the two interferometers can then be read out using the four output ports. The main plot shows gradiometer data. The two interferometers have a fixed phase difference independent of common mode phase noise. When plotted parametrically, the interferometer outputs form an ellipse whose shape is determined by this relative phase difference. Ellipses are plotted both with (red, hollow) and without (blue, filled) a laser beam applied to phase shift the lower interferometer by $\phi_{\text{ac}}$. For this data, the atoms separated for 63 ms, giving 1.764 mm of separation to the gradiometer. $T = \tau = 0.3$ ms was then used for the interferometers.
}
\label{gradiometer}
\end{figure}

These tools enable novel and flexible interferometer geometries. As an example, we realize a single-source gradiometer (Fig. \ref{gradiometer}a). A SDK beamsplitter is used to separate two arms of the atomic wavefunction. Once separated, they are brought back to equal velocity and used to perform two SDK interferometers simultaneously. These interferometers can then measure a relative phase, where common-mode noise (vibrations, laser phase noise) is rejected \cite{Foster2002MethodFitting}. We demonstrate the gradiometer in Fig. \ref{gradiometer}b by measuring a phase induced by a transverse laser beam incident on only the lower SDK interferometer. The upper and lower interferometers have the same velocity and the same internal states, reducing systematic effects. Additionally, the gradiometer baseline is known to high precision, since it is determined only by the photon momentum and wavepacket separation time.

\begin{figure}
\centering
\includegraphics[width = 3.38in]{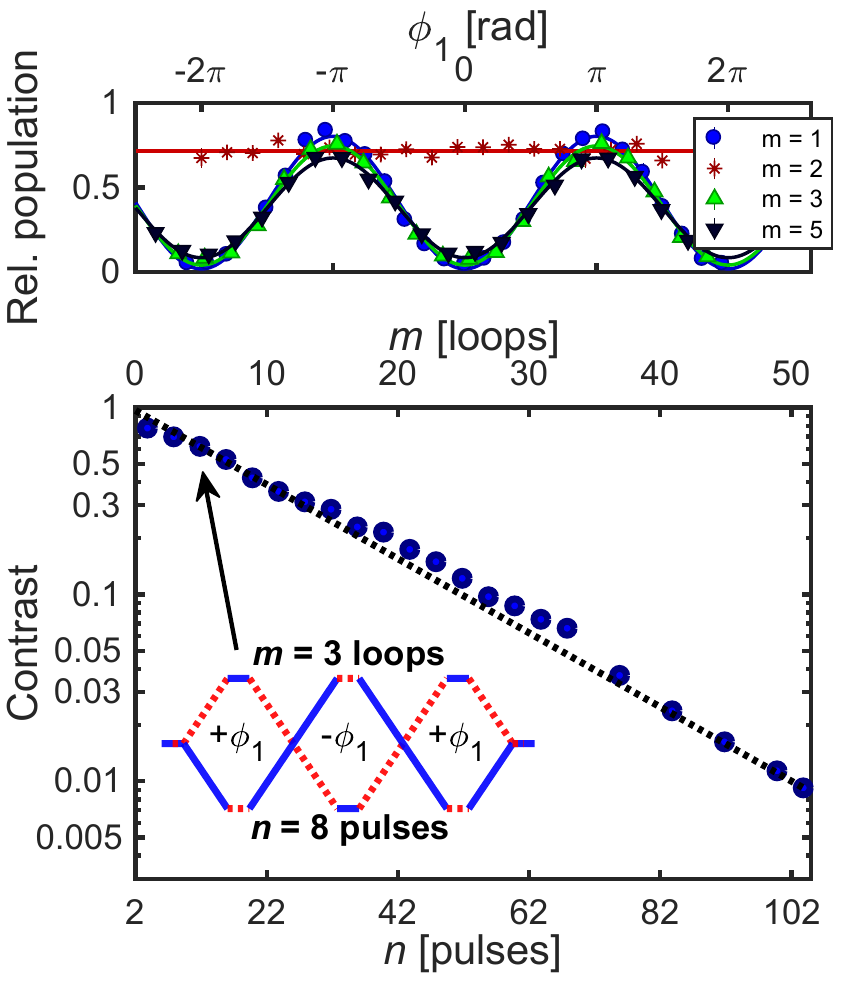} 
\caption{Resonant atom interferometer. Top: Interference fringes for different number of loops $m$, as the phase per loop is varied. Bottom: Contrast decay is shown as both a function of the number of loops $m$, and corresponding number of optical pulses $n$. Resonant interferometer geometry for $m = 3$ loops is illustrated in the lower left. The dotted line represents a model with no free parameters, using only the independently measured Ramsey contrast (88\%) and ARP pulse efficiency (96\%), and the calculated single photon scattering (1\% per pulse). Agreement with the data indicates negligible sources of additional contrast loss.
}
\label{juggling}
\end{figure}

As a further example, we demonstrate a tunable detector for ac accelerations. Such ``resonant" atom interferometers have been proposed to search for gravitational waves \cite{Graham2016ResonantInterferometry} or oscillating forces due to light dark matter \cite{Graham2016DarkAccelerometers}. As shown in Fig. \ref{juggling} (inset), lock-in ac detection is achieved by having the wave function enclose several loops ($m=3$ are shown). The sensitivity function reverses in each loop, as the arms are kicked in alternating directions. A requirement for such a detector is the efficient application of many pulses. Performing many loops increases the frequency selectivity (``quality factor" $Q$) of the resonant detector, and therefore its noise suppression at other frequencies. The frequency probed is set by the duration of each loop, which is easily tuned. 

We demonstrate a proof of principle for such a scheme. The top panel of Fig. \ref{juggling} confirms the expected behavior of such a resonant interferometer: for even $m$, dc effects (such as gravity and laser phase per loop $\phi_{1}$) cancel, and the interferometer phase remains zero regardless of $\phi_{1}$. For odd $m$ the net interferometer phase is that of a single loop, $\phi_{1}$. For this demonstration, contrast data was taken with loop sizes of $T = \tau = 10 \, \mu$s at $4\hbar k$ splitting to allow over 100 pulses of 200-$\mu$s duration to fit within the available free-fall time. A stable fringe is observed at each loop order, whose fitted amplitude matches the histogram-fitted contrast of Fig. \ref{juggling}. This indicates that phase noise per optical pulse is negligible. LMT could also be implemented in each loop to increase sensitivity.

All of the above schemes are insensitive to ac Stark shifts. In a typical Raman interferometer, only differences in the initial and final optical $\pi/2$ pulses contribute to ac Stark shifts.  We use only optical $\pi$ pulses where, roughly speaking, the atoms spend half of every pulse in each hyperfine state, cancelling the Stark shift \cite{Weiss1994PrecisionInterferometry}.  Indeed, testing an SDK interferometer by increasing the intensity of all pulses by up to a factor of two saw negligible effects. This is particularly advantageous in our interferometer, which operates in an optical cavity which complicates ac Stark shifts \cite{Jaffe2017TestingMass}.

We have demonstrated a new tool for light-pulse atom interferometers by splitting the beamsplitter into two operations. This simple change enables the exclusive use of highly efficient adiabatic passage, opening the door to a wide range of new and old geometries. The use of Raman atom optics and a thermal sample greatly relax the complexity required of the atom source to implement these geometries, without precluding their use in existing high-performance devices. This technique combines the advantages of Bragg transitions (LMT compatibility, ac Stark insensitivity) and Raman adiabatic passage (state-labeling, high efficiency, wide bandwidth).

This tool's flexibility allows specialization for multiple applications. Short pulses forming many loops near a source mass would constitute a lock-in force sensor probing viable mass ranges for light dark matter candidates \cite{Graham2016DarkAccelerometers}. High-power, large bandwidth pulses with fast, simple atom preparation could provide LMT for precise inertial sensing. Multi-pulse geometries, e.g., resonant AI or single-source gradiometer, enabled by high fidelities can provide technical benefits to existing and future measurements. A next step may envision a squeezed atom interferometer built using the collective cavity measurement demonstrated in \cite{Hosten2016MeasurementAtoms,Cox2016DeterministicFeedback}. We hope that SDK interferometry can make demanding experiments tractable, as well as improve sensitivity across a range of measurement types.

We thank Shannon Baucom for contributions to the apparatus and Wes Campbell for stimulating discussions. We thank Erik Urban and Andrew Eddins for useful discussions in resolving technical difficulties. This material is based upon work supported by grant number 2009-34712 by the David and Lucile Packard Foundation, numbers 1404566 and 1708160 by the National Science Foundation, as well as grants No. 1548805, 1568877, 
1570297, 1553641, 1584449, and 1584157 by the National Aeronautics and Space Administration. P. Haslinger thanks the Austrian Science Fund (FWF): J3680.  Both H. M{\"u}ller and P. Hamilton thank the UC Office of the President for grant CA-16-377655, and P. Hamilton thanks the Office of Naval Research for grant N000141712256.

\bibliographystyle{apsrev4-1}
\bibliography{sdk_arxiv_v0.bib}

\end{document}